\documentclass[aps,pre,twocolumn,superscriptaddress,showpacs]{revtex4}
\usepackage{amssymb}
\usepackage{epsfig}
\usepackage{amsmath}
\usepackage{times}
\usepackage{epstopdf}
\usepackage{amsfonts}

\begin{document}

\title{The relative importance of structure and dynamics on node influence in reversible spreading processes}

\author{Junyi Qu}
\affiliation{School of Physics and Electronic Science, East China Normal University, Shanghai 200241, China}

\author{Ming Tang}\email{mtang@ce.ecnu.edu.cn}
\affiliation{School of Physics and Electronic Science, East China Normal University, Shanghai 200241, China}
\affiliation{Shanghai Key Laboratory of Multidimensional Information Processing, East China Normal University, Shanghai 200241, China}

\author{Ying Liu} \email{shinningliu@163.com}\affiliation{School of Computer Science, Southwest Petroleum University, Chengdu 610500, China}

\author{Shuguang Guan}
\email{sgguan@phy.ecnu.edu.cn}\affiliation{School of Physics and Electronic Science, East China Normal University, Shanghai 200241, China}

\date{\today}

\begin{abstract}
The reversible spreading processes with repeated infection widely exist in nature and human society, such as gonorrhea propagation and meme spreading. Identifying influential spreaders is an important issue in the reversible spreading dynamics on complex networks, which has been given much attention. Except for structural centrality, the nodes' dynamical states play a significant role in their spreading influence in the reversible spreading processes. By integrating the number of outgoing edges and infection risks of node's neighbors into structural centrality, a new measure for identifying influential spreaders is articulated which considers the relative importance of structure and dynamics on node influence. The number of outgoing edges and infection risks of neighbors represent the positive effect of the local structural characteristic and the negative effect of the dynamical states of nodes in identifying influential spreaders, respectively. We find that an appropriate combination of these two characteristics can greatly improve the accuracy of the proposed measure in identifying the most influential spreaders. Notably, compared with the positive effect of the local structural characteristic, slightly weakening the negative effect of dynamical states of nodes can make the proposed measure play the best performance. Quantitatively understanding the relative importance of structure and dynamics on node influence provides a significant insight into identifying influential nodes in the reversible spreading processes.
\end{abstract}

\maketitle

\section{Introduction}
Identifying influential spreaders is an important topic in the spreading dynamics on complex networks \cite{koschutzki2005centrality,lu2016vital,pei2020influencer}. Identifying these nodes can help us to formulate precise marketing strategies \cite{leskovec2007dynamics}, control the spread of public opinions \cite{bovet2019influence,lin2018evolution}, prevent the catastrophic disruptions in power girds \cite{motter2002cascade,albert2004structural} and control the outbreak of epidemics \cite{pastor2002immunization,scarpino2019predictability,zhou2008epidemic}.

In recent years, there have been a lot of researches devoted to identifying the most influential spreaders \cite{morone2015influence,pei2018theories,hu2018local,lokhov2017optimal,zheng2021k}.
As the network structure has a significant impact on the spreading processes, some structural centralities are used to measure the spreading influence of a node or node set, such as degree, closeness centrality, betweenness centrality and $k$-shell index \cite{lu2016vital,pei2020influencer}.

However, the complex interplays between network structure and spreading dynamics sometimes make the structural centralities be unable to effectively identify the influential nodes accurately in complex networks. \cite{poux2020influential,erkol2020influence,aral2018social}.
Integrating the dynamical states of nodes in the spreading process can promote the accuracy of measures in identifying the most influential spreaders \cite{klemm2012measure}. Most of previous researches focus on irreversible spreading dynamics which has a final state. But, some spreading processes are reversible with repeated infections,
such as meme spreading on online social networks \cite{gleeson2014competition}, gonorrhea spreading propagation on sexual intercourse networks \cite{pastor2001epidemic} and the rise and fall of stock prices on financial stock market networks \cite{stavroglou2019hidden}. These reversible processes can be described by the SIS spreading dynamics \cite{barzel2013universality,pastor2001epidemicsp,pastor2001epidemicdynamics}, where a node transfers between infection state and recovery state.
Qu $et$ $al$. proposed a single-node control model to evaluate the influence of nodes and a structure-dynamics combined centrality to identify influential nodes in the reversible spreading system \cite{qu2020identifying}. They found that taking the neighbors' centrality and dynamical states into account can identify influential nodes more accurately in real-world networks than the benchmark centralities.

Although the interplay between network structure and dynamics determines the influence of nodes, the relative importance of the two factors in identifying influential nodes is still unknown. By considering the positive effects of the neighbors' local structures and the negative effects of the neighbors' dynamical states, a new structure-dynamics combined centrality is proposed to identify influential spreaders and evaluate the relative importance of structure and dynamics on the determining node influence. In this new measure, the number of outgoing edges and the infection risks of neighbors are used to represent the local structures and dynamical states of the neighbors respectively. Simulation results on real-world networks show that considering neighbors' local structures and dynamical states can help improving the accuracy of the index in identifying influential spreaders. Notably, compared with the number of outgoing edges of neighbors, slightly weakening the impact of neighbors' infection risks can greatly promote the accuracy of the index to identify influential spreaders. There is an optimal range, within which the local structures and dynamical states of neighbors enhance the accuracy of the influence ranking index the most.

The rest of the paper is organized as follows. Section II introduces a single-node control model to quantify the influence of a node in SIS spreading processes. In Section III, based on the neighbors' local structures and dynamical states , we propose a node centrality index in the SIS spreading dynamics. Section IV shows the numerical simulation results and analyses on real-world networks. Section V is the conclusion and discussion.

\section{Single-node control model in the SIS spreading processes}

In our previous research, we have proposed a single-node control method to evaluate the influence of nodes in the reversible spreading system \cite{qu2020identifying}. In SIS spreading processes, a node has only two possible states: susceptible (S) state and infected (I) state. An infected node transmits a disease to each of its direct susceptible neighbors with rate $\beta$ and recovers to susceptible state with rate $\mu$. After a long-time transient process, the proportion of infected nodes in the network will be stable, which is recorded as $\varphi_0$. Then we artificially set one node to be in the infected state constantly. This will cause a change (improvement) in the proportion of infected nodes in the entire network. After certain time steps, the proportion of the I-state nodes reaches a new steady state and the proportion of infected nodes is recorded as $\varphi$. We define $\Delta\varphi=\varphi-\varphi_0$ as the spreading influence, also called the spreading efficiency of the controlled node. The larger the $\Delta\varphi$ is, the higher influence of the controlled node has.

In simulations, we use the synchronous updating method \cite{shu2016recovery}. The recovery rate is $\mu=0.1$ and the effective transmission rate $\lambda=\beta/\mu$ should not be too large because when it is too large, the propagation can easily spread to the entire network regardless of the node under control \cite{liu2015core}. In that case the difference between the controlled nodes is diminished, and the influence of each node is indistinguishable. Meanwhile, the effective transmission rate $\lambda$ should not be too small. If it is too small, the propagation will be limited to the local neighbourhood of the controlled node and cannot affect the entire network. We obtain the SIS epidemic threshold $\lambda_c$ using the susceptibility \cite{ferreira2012epidemic,shu2015numerical,xu2019identifying}
\begin{equation}\label{yiganxing}
\chi(\lambda)=N\frac{<\rho^{2}>-<\rho>^{2}}{<\rho>},
\end{equation}
and choose an appropriate transmission rate $\lambda>\lambda_c$. Under this effective transmission rate, the influence of nodes can be distinguished.

\section{The proposed centrality NSRC}
It has been pointed out that compared with degree and $k$-shell index, considering the contribution of neighbors to the initial spreader can significantly improve the accuracy of ranking measures
in the SIR-like spreading processes \cite{liu2016identify}. If a neighbor $j$ of node $i$ only has neighbors that are also the direct neighbor of node $i$, when we control node $i$, the contribution of neighbor $j$ is less because it cannot assist node $i$ to spread the message to the entire network.

\begin{figure}
\centering
\includegraphics[width=0.8\columnwidth]{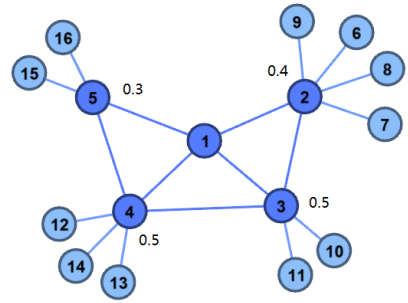}
\caption{Illustration of calculating the NSRC. The decimal number labeled to the node number is the probability of being in the infected state. The NSRC of node 1 is calculated as
$C(1)=k_1+a(k_2^{out}+k_3^{out}+k_4^{out}+k_5^{out})-b(k_2^{out}\rho_2+k_3^{out}\rho_3+k_4^{out}\rho_4+k_5^{out}\rho_5)=4+a(4+2+3+2)-b(1.6+1+1.5+0.6)=4+11a-4.7b$
}
\label{fig0}
\end{figure}

Assume that the network $G=(V,E)$ is unweighted and undirected composed of $|V|=N$ nodes and $|E|=M$ edges. The adjacency matrix of the network is denoted as $\textbf{A}=(a_{ij})$, where $a_{ij}=1$ means there is a direct connection between node $i$ and node $j$, otherwise $a_{ij}=0$. We further consider the two-step neighbor $l$ of node $i$ connected to the node $i$ through the nearest neighbour node $j$. If $a_{ij}=1$, $a_{jl}=1$ and $a_{il}\neq1$, then we define $b_{jl}=1$ to represent an outgoing edge of node $j$. The edge between the nearest neighbour node $j$ and node $l$ is the outgoing edge of node $j$ relevant to node $i$. For each of the neighbor node $j$, there is $k_j^{out}= \sum_{l\in\Gamma_j}{b_{jl}}$. Take the network in Fig. \ref{fig0} for example. Node 1 has four neighbors $\{2,3,4,5\}$ and the outgoing edges of its nearest neighbour node 2 are the edges that connect node 2 and the node set $\{6,7,8,9\}$.
The outgoing edges of nodes 3, 4, 5 connect to the node sets $\{10,11\}$, $\{12,13,14\}$ and $\{15,16\}$ respectively.
Considering the characteristics of SIS spreading processes, we introduce the number of outgoing edges of neighbors into the node influential index with a weighting parameter $a$. The idea is the more outgoing edges the neighbor nodes have, the more they help the considered node to spread information or message to the entire network \cite{liu2016identify}.

In the SIS spreading dynamics, when the system reaches a steady state, a node can still be infected repeatedly. In order to maximize the spreading to the entire network, we consider the risk that a node is infected.
In Fig. \ref{fig0}, the infection risk of node 3 is 0.5. When node 1 is controlled and spreads out through node 3, because node 3 has a risk of infection, the influence of node 1 will be reduced. As the infection risk has a negative impact on the influence of origin node, the term with $\rho$ has a minus sign. Parameter $b$ is an adjustable weight indicating the relative importance of local dynamical states (i.e., infection risks).

As considering neighbors' local structures and dynamical states can effectively improve the accuracy of ranking measure for node influence on them \cite{qu2020identifying}, we here investigate how to quantitatively evaluate the importance of these two factors. We propose an index to rank node influence
in SIS reversible spreading processes called neighboring local structures and infection risks coupled centrality (NSRC), which is defined as

\begin{equation}\label{NSRC}
C(i)=k_i+a\sum_{j\in\Gamma_i}{k_j^{out}}-b\sum_{j\in\Gamma_i}{k_j^{out}\rho_j},
\end{equation}
where $k_i$ is the benchmark centrality (i. e., degree), $j\in\Gamma_i$ is the direct neighbor of node $i$. The second item on the right side of Eq. (\ref{NSRC}) represents the local structures of the neighbor node $j$, which is the number of outgoing edges through all neighbors of node $i$. The third item represents the contribution of dynamical state of neighbor node $j$, which is the infection risk of node $j$. In the spreading process, when the system reaches its steady state, a node can still be infected repeatedly. In order to maximize the spread of infection to the entire network, we consider controlling a node $i$ in I state. Suppose the neighboring node $j$ is in the I state with infection probability $\rho_j$ in steady state. In simulations, if the neighboring node $j$ is in I state at this time step, node $i$ will have no effect on node $j$ and the influence of node $i$ will be reduced. As the $\rho_j$ has a negative impact on the influence of node $i$, the term with $\rho_j$ has a minus sign.
For simplicity, only the one-step neighbors of node $i$ and their outgoing edges are considered. $a$ and $b$ are two weight parameters, which represent contribution of the the local structures and the dynamical states of the neighbors respectively. The larger $a$ indicates the higher importance of the local structures and the larger $b$ indicates that the neighbors' infections are more important. The combination of $a$ and $b$ can be coordinated by their weights which will be discussed in detail later.

In order to measure the accuracy of NSRC in identifying influential spreaders, the following two evaluation methods are used. The imprecision function is defined as

\begin{equation}\label{epsilon}
\varepsilon(p)=1-\frac{\phi(p)}{\phi_{eff}(p)},
\end{equation}
where $p$ is the fraction of network size $N$. $\phi(p)$ is the average spreading efficiency of $pN$ nodes with the highest NSRC, and $\phi_{eff}(p)$ is the average spreading efficiency of the $pN$ nodes with the highest spreading efficiency. When $p$ is fixed, a smaller $\varepsilon(p)$ value indicates a more accurate centrality in identifying influential spreaders.

The imprecision function reflects the accuracy of the NSRC in identifying the most influential nodes. In order to  measure the ability of NSRC in ranking all nodes in the network, the Kendall's $\tau$ correlation coefficient is used, which is given by \cite{kenda1938new}
\begin{equation}\label{eq:eps4}
\tau=\frac{\sum_{i<j} \operatorname{sgn}[(x_i-x_j)(y_i-y_j)]}{\frac{1}{2}N(N-1)},
\end{equation}
where $\operatorname{sgn}(x)$ is a sign function. $\operatorname{sgn}(x)$ returns 1 if $x>0$, -1 if $x<0$, and 0 if $x=0$, respectively. $N$ is the number of nodes in the list. $x_i$ and $x_j$ are the ranking orders of node $i$ and node $j$ in sorting sequence 1. $y_i$ and $y_j$ are the ranking orders of node $i$ and node $j$ in sorting sequence 2. If node $i$ and node $j$ have a concordant order in sorting sequence 1 and 2, $(x_i-x_j)(y_i-y_j)>0$. If node $i$ and node $j$ have a discordant order in sorting sequence 1 and 2, $(x_i-x_j)(y_i-y_j)<0$. If node $i$ and node $j$ have a same order in sorting sequence 1 or 2, $(x_i-x_j)(y_i-y_j)=0$.
We rank the nodes by the proposed NSRC as the sorting sequence 1 and rank the nodes by the simulated spreading efficiency as sorting sequence 2. We compute the correlation coefficient of sorting sequence 1 and 2. The larger correlation coefficient is, the more concordant the centrality and the spreading efficiency are. For an explicit comparison, we calculate the improved $\tau$ ratio of the NSRC over the reference centrality, which is
\begin{equation}
\label{eta}
\eta=\left\{
\begin{aligned}
\frac{\tau_{c}-\tau_{0}}{\tau_{0}}      &      & {\tau_0>0};\\
\frac{\tau_{c}-\tau_{0}}{-\tau_{0}}     &      & {\tau_0<0};\\
0     &      & {\tau_0=0},
\end{aligned}
\right.
\end{equation}
where $\tau_c$ is the correlation coefficient between the NSRC and spreading efficiency, $\tau_0$ is the correlation coefficient between the neighboring dynamical information centrality (NDIC) and spreading efficiency.
\section{Numerical simulations}
We apply the NSRC to eight real-world networks as listed in Table \ref{network}. The real-world networks are: Netsci (collaboration network of network scientists), Hamster (friendships and family network of a website), Router (the router network collected by the Rocketfuel Project), Blog (communication network between users on the MSN website), PGP (an encrypted communication network), CA\_hep (collaboration network of arxiv in high-energy physics theory), Astro (collaboration network of astrophysics scientists) and Email (email communication network of University College London) \cite{newman2006finding,spring2002measuring,boguna2004models,newman2001the,kitsak2010identification}.
\begin{table*}
\centering
\renewcommand {\thetable} {\arabic{table}}
\begin{tabular}{lccccccccc}

\hline
Network& $N$    & $E$        & $<k>$   &$k_{max}$  &$H_k$     &$r$    &$C$    &$\lambda_c$ &$\lambda$ \\
\hline
Net379  &  379   &   914      & 4.8    &  34     &  1.663    &  -0.082  & 0.741  &  0.20   &0.28   \\
Hamster &  2000  &   16097    & 16.1   &  273    &  2.719    &  0.023   & 0.540  &  0.024  &0.04   \\
Router  &  5022  &   6258     & 2.5    &  16     &  5.503    &  -0.138  & 0.012  &  0.10   &0.20   \\
Blog    &  3982  &   6803     & 3.4    &  189    &  4.038    &  -0.133  & 0.284  &  0.10   &0.20   \\
PGP     &  10680 &   24340    & 4.6    &  206    &  4.153    &  0.240   & 0.226  &  0.06   &0.12   \\
CA\_Hep &  8638  &   24806    & 5.7    &  65     &  2.261    &  0.239   & 0.482  &  0.08   &014    \\
Astro   &  14845 &   119625   & 16.1   &  360    &  2.820    &  0.228   & 0.670  &  0.02   &0.10   \\
Email   &  12625 &   20362    & 3.2    &  576    &  34.249   &  -0.387  & 0.109  &  0.01   &0.06   \\
 \hline
\end{tabular}
\caption {Characteristics of the real-world networks studied in this paper. Theses characteristics include the number of nodes ($N$), number of edges ($E$), average degree ($<k>$), maximum degree ($k_{max}$), degree heterogeneity ($H_k$), degree assortativity ($r$), clustering coefficient($C$), epidemic threshold ($\lambda_c$) and disease-transmission rate $\lambda$ used in the SIS spreading processes.
}
\label{network}
\end{table*}
\subsection{The relative importance of neighbors' local structures and dynamical states}
\begin{figure*}
      \centering
      \includegraphics[width=0.9\textwidth]{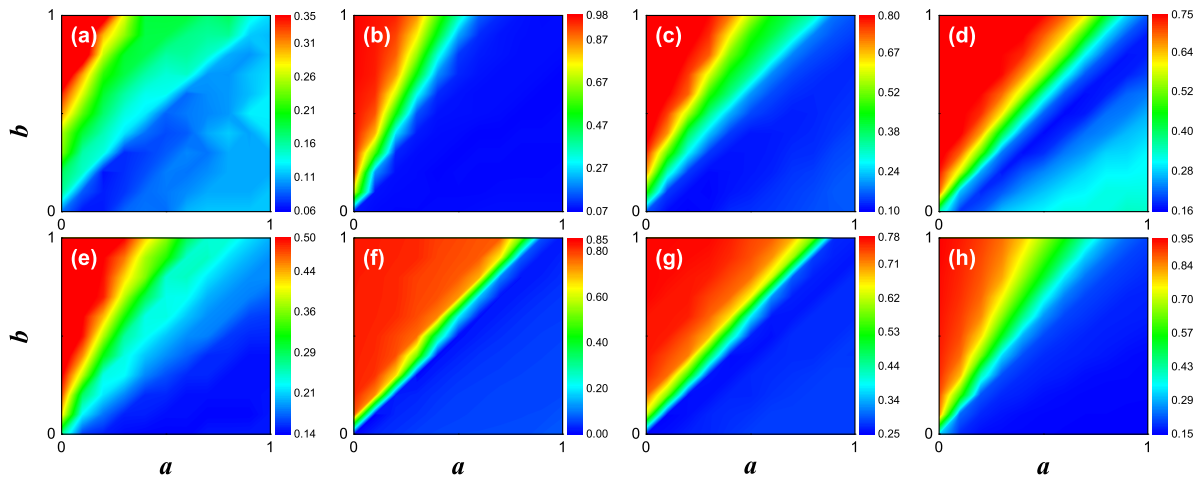}\\
      \caption{The imprecision of NSRC as a function of $a$ and $b$ in eight real-world networks for $p=5\%$. The real-world networks are Nstsci (a), Hamster (b), Router (c), Blog (d), CA\_Hep (e), Email (f), PGP (g), Astro (h).}
      \label{fig1}
\end{figure*}

\begin{figure*}
      \centering
      \includegraphics[width=0.9\textwidth]{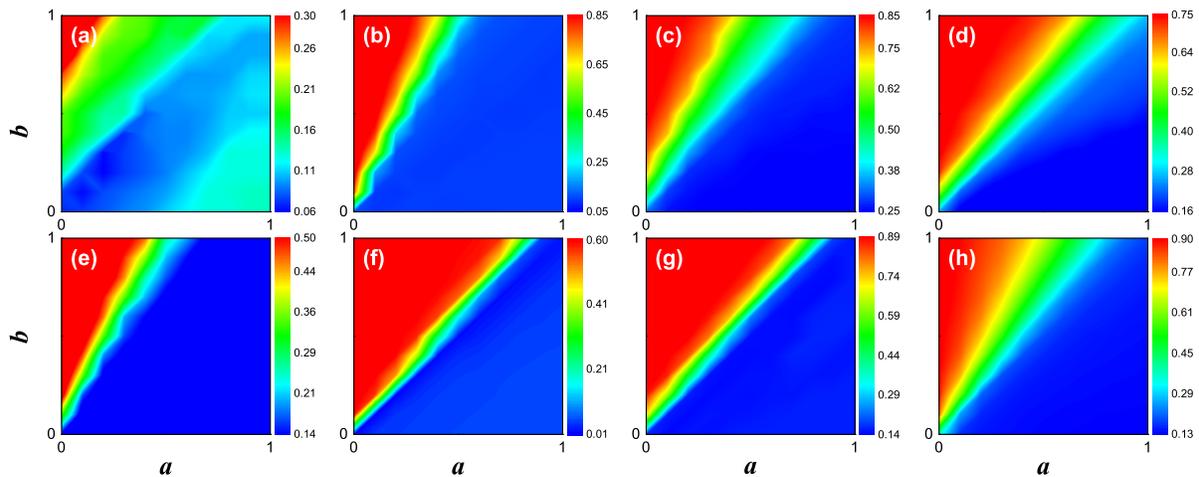}\\
      \caption{The imprecision of NSRC as a function of $a$ and $b$ in eight real-world networks for $p=20\%$. The real-world networks are Nstsci (a), Hamster (b), Router (c), Blog (d), CA\_Hep (e), Email (f), PGP (g), Astro (h).}
      \label{fig2}
\end{figure*}

Next we focus on the most influential spreaders which are very critical in the SIS spreading processes. We investigate the relative importance of the neighbors' local structures and dynamical states based on the top 5\% nodes and top 20\% nodes ranked by the NSRC. Figures \ref{fig1} and \ref{fig2} show the the imprecisions of the NSRC to identify the top 5\% and 20\% nodes under different values of the weight parameters $a$ and $b$ respectively. In the Fig. \ref{fig1}, where the top 5\% influential spreaders are considered, and the blue areas in which the imprecision of NSRC is relatively small correspond to the parameter range $a\geq b$ in the lower right part of the diagram. In the Netsci, Router, Blog, Email and PGP networks respectively, there is an optimal area at the lower part of the diagonal $a=b$. In the Hamster, CA\_Hep and Astro networks respectively, the imprecision is relatively small in the area of $a<b$. There is a saturation effect in this area where the imprecision does not change much with the change of $a$ and $b$. When we consider the top ranked 20\% nodes by the NSRC, the results shown in Fig. \ref{fig2} are similar to that of the top ranked 5\% nodes shown in Fig. \ref{fig1}. In the Netsci and Email networks, there are optimal areas near the diagonal while in the other six networks, the optimal area is the lower right part of $a>b$. This implies that for the proposed index NSRC based on neighbors' local structures and dynamical states to be more accurate, the weight of the infection risks $b$ should not exceed the weight of the neighbors' local structures $a$.


\begin{figure*}
      \centering
      \includegraphics[width=0.9\textwidth]{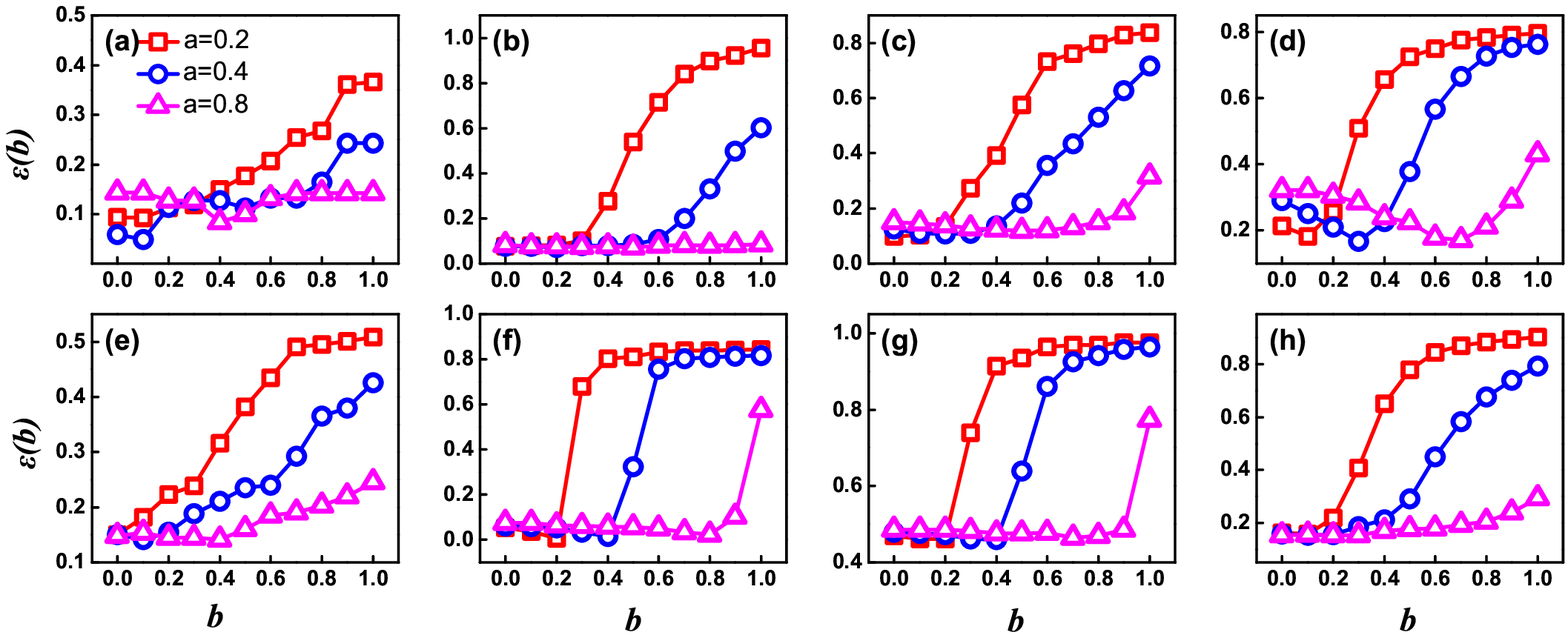}\\
      \caption{The imprecision of NSRC as a function of $b$ in eight real-world networks for $p=5\%$. The real-world networks are Nstsci (a), Hamster (b), Router (c), Blog (d), CA\_Hep (e), Email (f), PGP (g), Astro (h).}
      \label{fig3}
\end{figure*}

To further explore the impact of the local structures and dynamical states of neighbor nodes on the accuracy of the proposed centrality, we fix the value of the parameter $a$ and investigate the imprecisions of NSRC as a function of the parameter $b$. Figure \ref{fig3} shows that when the weight parameter $a$ of the neighbors' local structure is fixed, the imprecision function has an optimal value or saturation effect with the change of $b$. When $a=0.2$, in the Blog, Email and PGP networks, the imprecisions are lowest at around $b=0.1,b=0.2,b=0.2$ respectively. In the other five networks, the imprecisions increase with the increase of $b$, and there is a saturation effect when $b\leq a=0.2$. The results for $a=0.4$ and $a=0.8$ support the same conclusion, as shown in Fig. \ref{fig3}. These results indicate that the same conclusion also applies when $a=0.4$ and $a=0.8$ in Fig. \ref{fig3}. These results indicate that the parameters $a$ and $b$ have a synergistic effect in determining the accuracy of the centrality based on the neighbors' local structures and dynamical states. When the accuracy is the highest, the optimal parameter $b$ changes with the parameter $a$.

Through the experimental results in eight real-world networks, it is found that regardless of the value of the parameter $a$, the imprecision function will have a minimum value if $b$ is not greater than $a$. To understand the above phenomenon, we can rewrite the centrality index as

\begin{equation}\label{NSRC}
C(i)=k_i+a\sum_{j\in\Gamma_i}{k_j^{out}(1-\frac{b}{a}\rho_j)}.
\end{equation}
In the form of Eq. (\ref{NSRC}), we can discuss the relationship between the items more conveniently. The optimal area of $b<a$ indicates that although both the positive effect of the outgoing edges and the negative effect of the infection risks of neighbors' have a significant contribution to the accuracy of the ranking measure, the relative importance of them determines the ranking accuracy. Specifically speaking, compared to the positive effect of the number of neighbors' outgoing edges, slightly weakening the negative effect of neighbors' infection risks can get better identification results.

The above results can be explained as follows. For a considered node, if two of its neighbors have the same number of outgoing edges, a higher infection risk of its neighbor $\rho_j$ means a limited impact towards the spreading processes. However, a larger $\rho_j$ often implies that node $j$ tends to be closely connected to the considered node and its neighbor set (i.e., node i and its neighbor node $j$ having more common neighbors) \cite{boguna2013nature,castellano2012competing}. For example, as there are more common neighbors between node 1 and its neighbor node 3, the node 3 has a higher infection risk (i. e., infection probability) than neighbor node 5.
When the considered node 1 is controlled as an I-state node, the neighbor node with more common neighbors can enhance the influence of this neighbor node's outgoing propagation to a certain extent due to the echo chamber effect of local close structures \cite{zhang2014suppression,chen2018suppressing,wang2015dynamics}. Combining the above two effects, an appropriate parameter combination of $b$ being slightly less than $a$ can effectively and quantitatively distinguish the impact of neighbors with different infection risks. For example, for node 3 and node 5 in Fig. \ref{fig0}, under the echo chamber effect, the infection risk increases from 0.5 to 0.8, and 0.3 to 0.65, respectively. When $b=0$, the infection risks of these two nodes is not considered. Since the number of outgoing edges is the same, the two nodes have the same contribution to the spreading capacity of the controlled node 1. When $b=1$, although the infection risk of node 5 is significantly greater than that of node 3, the negative effect of high infection risk towards the spreading range of the node 1 is overvalued due to the enhancement of the echo chamber effect of local close structures. But in fact we need to consider the impact of the echo chamber effect and appropriately decrease the weight of the neighbor's infection risk, thus the value of $b$ should be less than $a$.

\subsection{Compared with NDIC}
In our previous study, we proposed a similar ranking index based on neighboring dynamical information (NDIC) \cite{qu2020identifying}. Compared with the NDIC, the newly proposed NSRC classifies the neighbors' edges and especially addresses the impact of outgoing edges.
For a better illustration, we compared the two indicators by the Kendall's $\tau$ correlation and calculated the improved $\tau$ ratio $\eta$ of the new index over the existing NDIC.
It has been discussed in Ref. \cite{qu2020identifying} that when $a=0.2$, the accuracy of NDIC is the highest or a saturation effect appears when $b=0.5$.

\begin{figure*}[htb]
      \centering
      \includegraphics[width=0.9\textwidth]{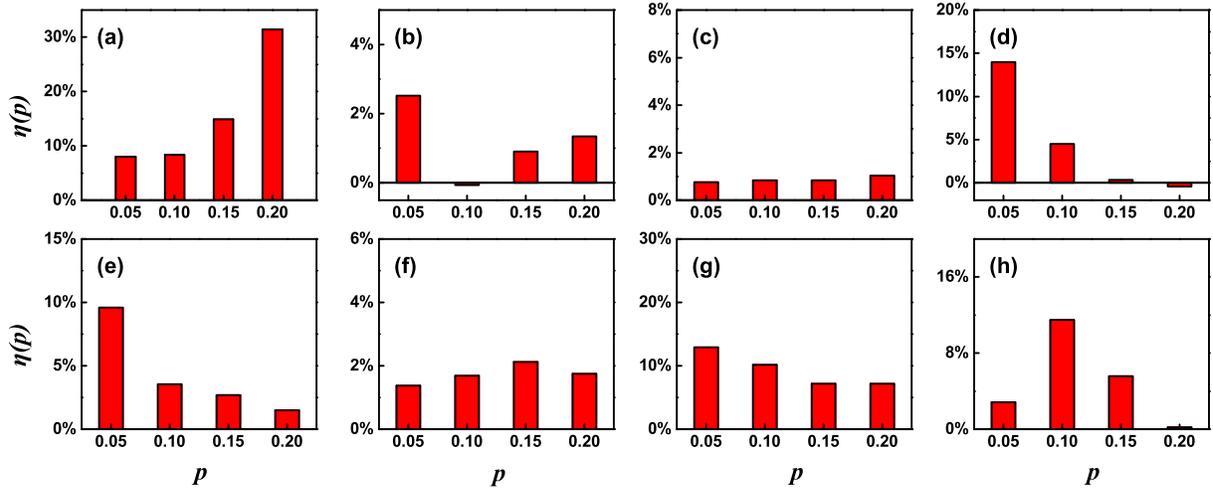}\\
      \caption{The improved ratio $\eta$ of NSRC over NDIC as a function of $p$. The weight parameters of $a$ and $b$ are both $a=0.2,b=0.5$. The real-world networks are Nstsci (a), Hamster (b), Router (c), Blog (d), CA\_Hep (e), Email (f), PGP (g), Astro (h).}
      \label{fig4}
\end{figure*}

\begin{figure*}
      \centering
      \includegraphics[width=0.9\textwidth]{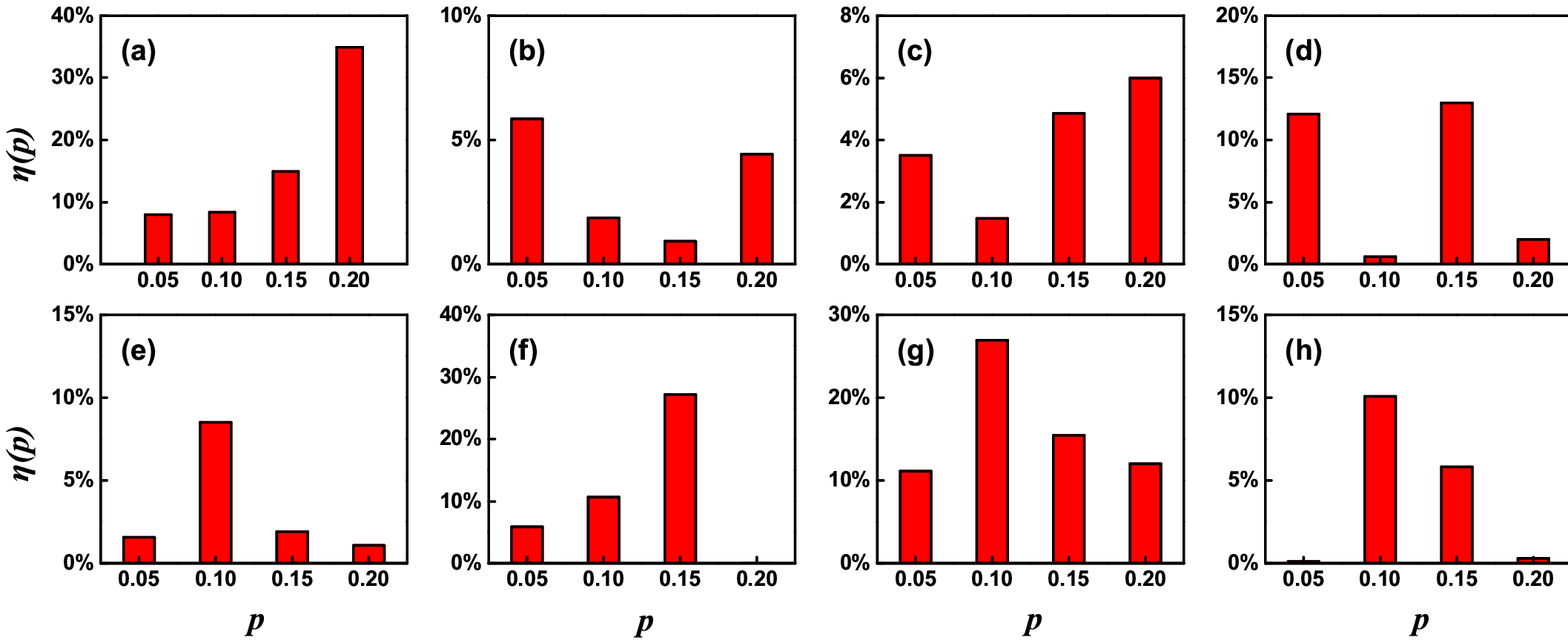}\\
      \caption{The improved ratio $\eta$ of NSRC over NDIC as a function of $p$. The weight parameters of $a$ and $b$ for NSRC and NDIC are $a=0.2,b=0.1$ and $a=0.2,b=0.5$ respectively. The real-world networks are Nstsci (a), Hamster (b), Router (c), Blog (d), CA\_Hep (e), Email (f), PGP (g), Astro (h). }
      \label{fig5}
\end{figure*}

\begin{figure*}
      \centering
      \includegraphics[width=0.9\textwidth]{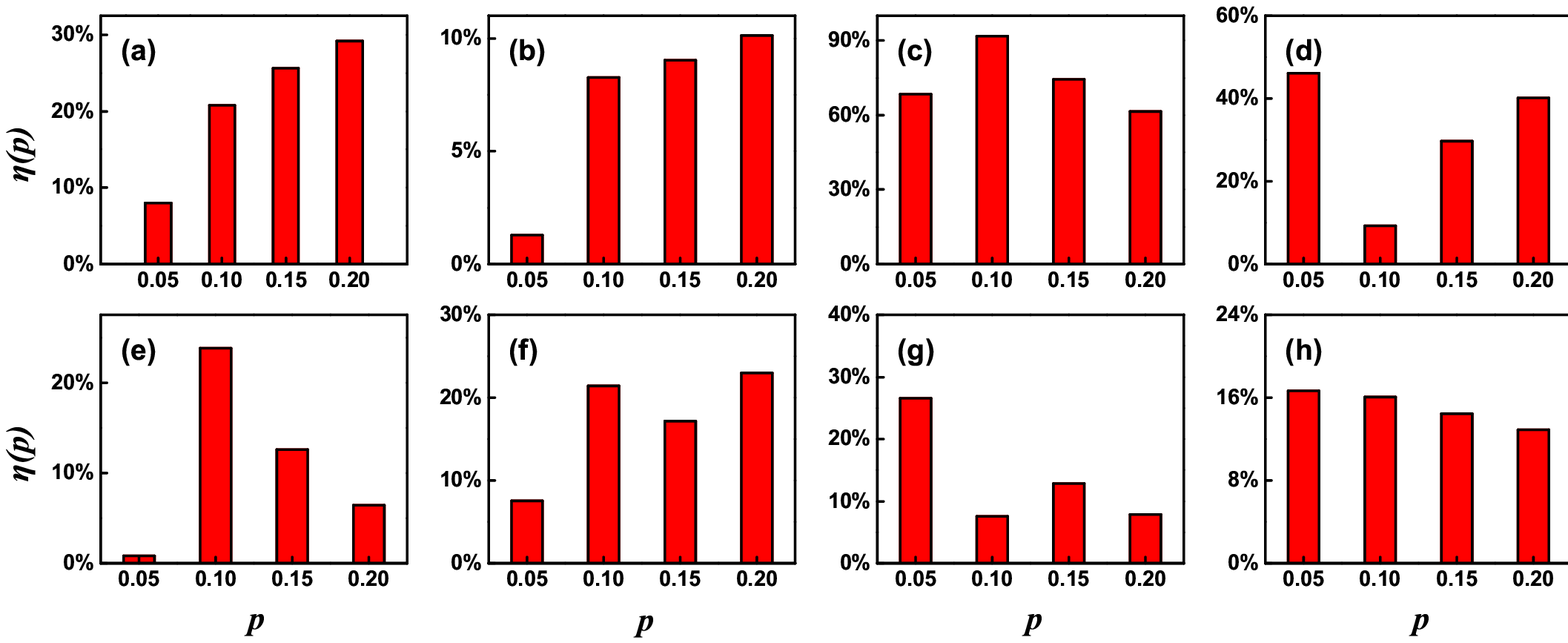}\\
      \caption{The improved ratio $\eta$ of NSRC over NDIC as a function of $p$. The weight parameters of $a$ and $b$ for NSRC and NDIC are $a=0.4,b=0.2$ and $a=0.2,b=0.5$ respectively. The real-world networks are Nstsci (a), Hamster (b), Router (c), Blog (d), CA\_Hep (e), Email (f), PGP (g), Astro (h).}
      \label{fig6}
\end{figure*}

In this combination of $a$ and $b$, the increased correlation ratio $\eta$ of NSRC over NDIC is greater than 0 in most cases as shown in Fig. \ref{fig4}. We also see that the NSRC displays a greater increase in ranking accuracy than the NDIC in the all networks, when an appropriate combination of parameters is fixed, as shown in Fig. \ref{fig5}. If we choose an optimal parameter combination of $a=0.4$ and $b=0.2$, the improved $\tau$ ratio of NSRC over NDIC is much more significant as shown in Fig. \ref{fig6}.
In the Router network, the improved $\tau$ ratio $\eta$ is as high as 90\%. In the Blog network, $\eta$ is close to 50\%. In the other six networks, $\eta$ is up to 10\% to 30\%.

\begin{figure*}
      \centering
      \includegraphics[width=0.9\textwidth]{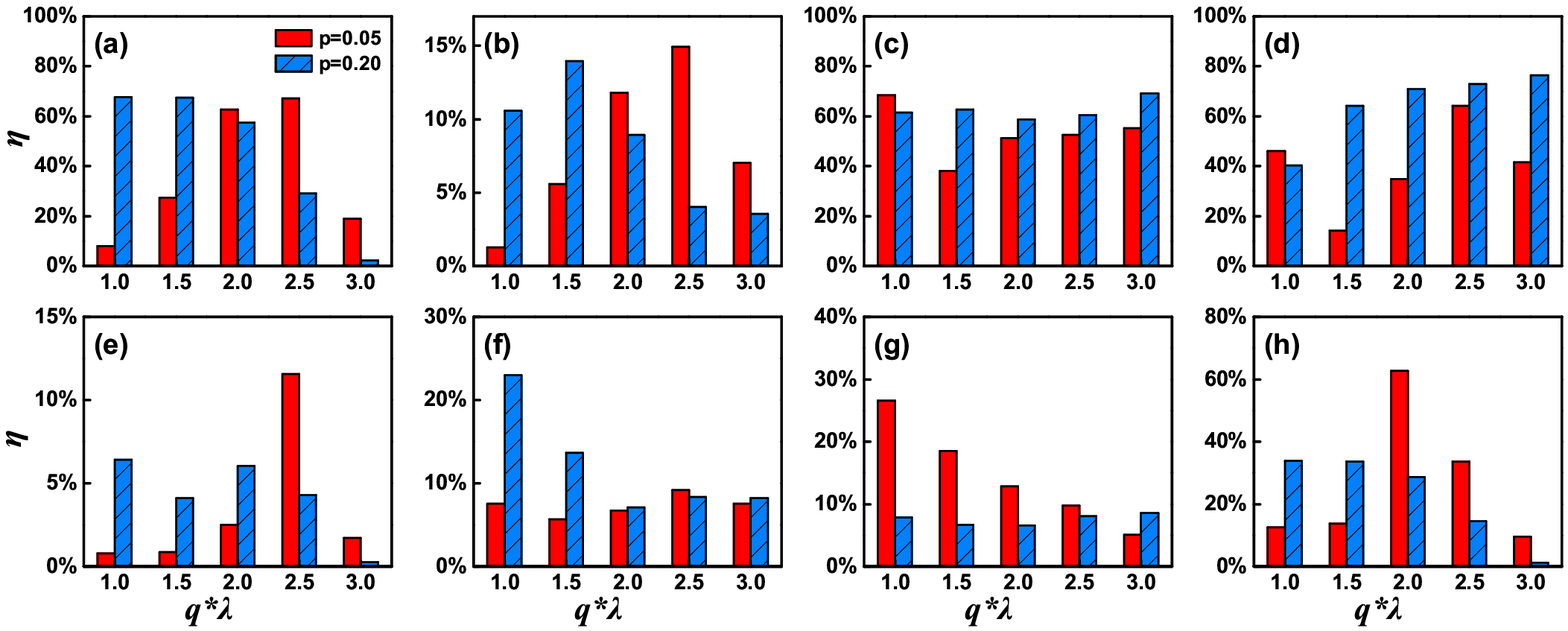}\\
      \caption{The improved ratio $\eta$ of NSRC over NDIC as a function of $\lambda$ in eight real-world networks. The imprecisions for $p=5\%$(red column) and $20\%$(blue column) are compared in each network. The real-world networks are Nstsci (a), Hamster (b), Router (c), Blog (d), CA\_Hep (e), Email (f), PGP (g), Astro (h).}
      \label{fig7}
\end{figure*}

Next, we investigate the dependence of $\eta$ on the transmission rate $\lambda$. The transmission rate should not be too large when distinguishing the node influence by using the single-node control model. Because when $\lambda$ is too large, a node with low centrality can also spread the disease to the entire network. In this case the influence of nodes cannot be distinguished. We thus use the 1 to 3 times of the epidemic threshold as the transmission rate \cite{liu2015core}. The improved ratio $\eta$ of NSRC over NDIC as a function of transmission rate $\lambda$ in eight real-world networks are shown in Fig. \ref{fig7}. The weight parameters are set to be $a=0.4,b=0.2$ and $a=0.2,b=0.5$ for NSRC and NDIC respectively. In the Netsci, Router, and Blog networks, the improved $\tau$ ratio is about 80\%. In the rest five networks, $\tau$ is around 10\% to 30\%. These results indicate that the ranking accuracy of NSRC is better than that of NDIC at various transmission rates.

\section{Conclusions}
In this paper, we quantitatively analyzed the relative importance of the neighbors' local structures and dynamical states in ranking node influence in the SIS-like spreading processes. We proposed a centrality index NSRC that combines the positive effect of the neighbors' outgoing edges and the negative effect of the neighbors' infection risks with two weighting factors $a$ and $b$. A large number of neighbors' outgoing edges have a positive effect on the spreading influence of a node, while the high infection risks of the neighbor nodes reduce the influence of the considered node. A larger value of $a$ or $b$ means a stronger positive local structure effect or negative local dynamical effect. Through a large number of simulations on eight real-world networks, we found that when the positive and negative effects are taken appropriately, under appropriate values of $a/b$, the proposed centrality index NSRC can improve the ranking accuracy significantly.

Specifically, to identify the most influential nodes accurately, the weight of the neighbors' dynamical states $b$ should not be greater than the weight of the neighbors' local structures $a$. Although the high infection risk of a neighbor node can inhibit the spreading capacity of the controlled node, the close local structure with more common neighbors between the considered node and its neighbor node can induce the echo chamber effect, and thus slightly promote the spreading processes. Therefore, an optimal combination of parameters with $b<a$ can make the NSRC evaluate the spreading influence of a node more accurately. This is helpful to distinguish the impact of the neighboring nodes having the same local structures but different infection probabilities. By analyzing the simulation results of real-world networks, it is found that when the outgoing edges of neighbor nodes are taken into account, the accuracy of NSRC is greatly improved over the centrality measure NDIC. After further optimizing the parameters, the improved ranking correlation of NSRC over NDIC can be as high as 10\% to 100\%. Compared with the degree centrality of the neighbors', the neighbors' outgoing edges can be more accurate in reflecting the importance of neighbors' local structures in the spreading processes. Further more, when we change the transmission rate $\lambda$, the NSRC has a higher accuracy than the existing index in ranking influential spreaders in the SIS-like spreading processes.

In the reversible processes, we found that considering the positive effects of neighbors' outgoing edges and the negative effects of neighbors' infection risks can effectively improve the accuracy of indexes in identifying influential spreaders. This deepens our understanding of identifying influential spreaders in the reversible processes. The benchmark centrality used in NSRC in our paper is degree, which can be substituted by any other benchmark centralities, such as $k$-shell index centrality \cite{kitsak2010identification} and the collective influence index \cite{morone2015influence}.
The proposed ranking index NSRC can be applied in other reversible dynamics and dynamical systems with steady state such as synchronization processes \cite{barzel2013universality} and cascading failures \cite{lin2020non}.

\acknowledgments

This work was supported by the National Natural Science Foundation of China (Grant Nos. 11875132, 11975099, 82161148012, 11835003, 61802321), the Natural Science Foundation of Shanghai (Grant No. 18ZR1411800), and the Science and Technology Commission of Shanghai Municipality (Grant No. 14DZ2260800).

\bibliography{ref}

\end{document}